\documentclass[10pt,aps,prc,floatfix,twocolumn,nofootinbib,superscriptaddress]{revtex4-1}
\usepackage{graphicx,amsmath,amssymb,bm}
\usepackage{amsfonts}
\usepackage[utf8]{inputenc}
\usepackage{verbatim}
\usepackage{float}
\usepackage[labelfont={small},subrefformat=parens,caption=false]{subfig}
\usepackage{cancel}
\usepackage{multirow}
\usepackage{array}
\usepackage{xparse}
\usepackage{xspace}

\usepackage{bigstrut}

\usepackage{physics}
\usepackage{color}
\usepackage{dsfont}

\newcommand{\beq}{\begin{equation}}
\newcommand{\eeq}{\end{equation}}
\newcommand{\bseq}{\begin{subequations}}
\newcommand{\eseq}{\end{subequations}}
\newcommand{\bi}{\begin{itemize}}
\newcommand{\ei}{\end{itemize}}
\newcommand{\be}{\begin{enumerate}}
\newcommand{\ee}{\end{enumerate}}

\newcommand{\Vopt}{V_{\text{opt}}}

\newcommand{\doublederivativeX}{\frac{d^2}{dx^2}}
\newcommand{\Kinetic}{\frac{p^2}{2m}}

\newcommand{\psiGround}{\psi_g(x)}
\newcommand{\psiExcited}{\psi_e(x)}

\newcommand{\kGround}{k_g}
\newcommand{\kExcited}{k_e}

\newcommand{\ts}{\textstyle}

\newcommand{\ds}{\displaystyle}
\newcommand{\wt}{\widetilde}

\newcommand{\Gtilde}{\widetilde G}
\newcommand{\Gzeroe}{G^0_e}
\newcommand{\Gzerog}{G^0_g}

\usepackage{dcolumn}

\usepackage[pdfencoding=auto, pdfpagelabels]{hyperref}

\usepackage[overload]{textcase}

\definecolor{linkcolor}{rgb}{0,0,0.40} 
\hypersetup{%
    pdfsubject=Paper,
    pdfkeywords={nuclear physics} {Bayesian} {chiral EFT} {emulator},
    unicode = true,
    breaklinks = true,
    colorlinks = true,
    linkcolor = linkcolor,
    citecolor = linkcolor,
    menucolor = linkcolor,
    urlcolor = linkcolor
}

\usepackage{cellspace}
\graphicspath{{./figures/}}

\newcommand{\pdag}{\vphantom{\dagger}}

\begin{document}

\title{Renormalization group evolution of optical potentials: explorations using a toy model}

\author{M.~A.  Hisham}
\email{hisham.3@osu.edu}
\affiliation{Department of Physics, The Ohio State University, Columbus, OH 43210, USA}

\author{R.~J. Furnstahl}
\email{furnstahl.1@osu.edu}
\affiliation{Department of Physics, The Ohio State University, Columbus, OH 43210, USA}

\author{A.~J. Tropiano}
\email{tropiano.4@osu.edu}
\affiliation{Department of Physics, The Ohio State University, Columbus, OH 43210, USA}

\date{\today}

\begin{abstract}
To take full advantage of experimental facilities such as FRIB for applications to nuclear astrophysics, nuclear structure, and explorations of neutrinos and fundamental symmetries, we need a better understanding of the interplay of reaction and structure theory.
The renormalization group (RG) is the natural tool for maintaining a consistent treatment of reaction and structure.
Here we make a first study of RG for optical potentials, which are important ingredients for direct reactions.
To simplify the analysis, we use a pedagogical one-dimensional model and evolve toward low RG resolution using the similarity RG (or SRG). 
We show how SRG decoupling at low resolution carries over to the optical potential and enhances perturbative approximations, and how induced SRG nonlocality compares to the nonlocality of the optical potential.
We discuss the results in the larger context of consistent SRG evolution of operators and wave functions in the analysis of direct reactions.
\end{abstract}

\maketitle

\section{Introduction} \label{sec:intro}

Experiments at the Facility for Rare Isotope Beams (FRIB) and other laboratories offer the potential for new insight into nuclear astrophysics, nuclear structure, and the physics of neutrinos and fundamental symmetries~\cite{LRP:2015}. 
To realize this potential we need to better understand the analysis of direct reactions.
In particular, we seek to isolate quantities of interest in experiments in a controlled manner, which implies process independent analyses, e.g., to apply measurements from one process to another process or to use them to cleanly extract structure information.

A specific challenge is that the separation of reaction and structure is not unique, even though many experimental analyses assume implicitly that it is. 
This non-uniqueness can be swept under the rug as ``model dependence'', but we can do better by working in a renormalization group (RG) framework. 
We will focus on the Similarity RG (or SRG)~\cite{Bogner:2006pc}, although there are other choices for such a framework~\cite{Bogner:2009bt}.
Under SRG evolution, the dividing line between structure and reaction, which is a momentum scale we call the RG resolution (in practice the largest momentum component of low-energy wave functions), shifts continuously.
Structure and reaction elements individually depend on this RG scale (``scale dependence'') but when put together the observable quantities are invariant, due to the evolution being a series of continuous unitary transformations.
There is also individual dependence of these elements on the starting Hamiltonian and the details of the RG implementation (``scheme dependence''), but observables are again unchanged during the evolution.

An RG framework provides control over scale and scheme dependence and enables us to relate theory calculations at different scales~\cite{More:2015tpa,More:2017syr}.
In Refs.~\cite{Tropiano:2021qgf,Tropiano:2022jjj}, the necessity and technical details of treating reactions and structure at the same scale were demonstrated explicitly for the SRG.
These works in particular illustrated the advantages of analyzing experiments targeted at short-range correlations (SRCs) at a low RG resolution (not to be confused with the experimental resolution), namely one that is most natural for nuclear ground and low-lying states (e.g., the Fermi momentum). 
As part of a broader effort to generalize these studies,
in the present work we take the first steps toward investigating and understanding optical potentials in an SRG framework. 

Optical potentials are effective interactions that describe the propagation of a particle through a many-body target~\cite{Hodgson:1971ab,Dickhoff:2018wdd,Holt:2022piv}.
They are instrumental in reducing a many-body scattering problem to a few-body process, such as that between projectile-target or residue-target.
Feshbach's seminal work formalized the optical potential and described three key characteristcs:
it is complex, nonlocal, and energy dependent~\cite{Feshbach:1958nx}.
Subsequent treatments identified the optical potential with the single-particle self-energy, which enabled dispersion relation approaches~\cite{Mahaux1991,Dickhoff:2018wdd}. 

Significant progress has been made in ab initio nuclear reactions, in which structure and reaction is automatically treated consistently, with several approaches to ab initio optical potentials~\cite{Holt:2013tna,Vorabbi:2015nra,Rotureau:2016jpf,Gennari:2017yez,Furumoto:2019anr,Idini:2019hkq,Durant:2022bjb}.
However, the ingredients and implications need further study. 
For example, the impact on the accuracy of observables of scale and scheme dependence and the optical potential's truncation is not yet under control.  
Also, the approximations necessitated in ab initio approaches and the computational limitations to light nuclei means that phenomenological approaches are still needed.
A key question is then: how can we 
implement the needed structure-reaction consistency in a phenomenological (or semi-phenomenological) approach, so as to minimize ambiguities from scale and scheme dependence? 

There is a natural affinity between optical potentials and RG (and the SRG in particular) because they share the common feature of decoupling degrees of freedom. 
This connection has not yet been explored in the literature. 
In contrast, the effect of SRG evolution on Hamiltonians and (some) operators has been explored and exploited, both in free space and in the medium~\cite{Bogner:2006pc,Bogner:2009bt,Furnstahl:2013oba,Hergert:2016iju}. 
Evolution of Hamiltonians defined at high-momentum scales relative to the nuclear Fermi momentum (a natural resolution scale for nuclei) to lower resolution leads to softened potentials and an increased separation of scales advantageous for the treatment of interactions~\cite{Anderson:2010aq, More:2017syr, Tropiano:2020zwb, Tropiano:2021qgf}. 
Low RG resolution would seem to be implied by optical potential phenomenology because of the use of shell model wavefunctions~\cite{Tostevin:2014usa,Tostevin:2021ueu}. 
How can we make this consistent? 
Another open issue is the interplay of nonlocality from SRG decoupling of momentum modes with the nonlocality arising in optical potentials.

Analyzing the SRG evolution of an optical potential is a formidable challenge. 
It is not just a matter of evolving the (complex) optical potential by itself in an SRG equation as one would evolve the full potential (see discussion in Sec.~\ref{sec:summary}). 
Indeed, the most immediate path at present is
to solve the entire reaction problem at each SRG scale and then construct the optical potential at that scale.

We have faced such challenges before in the development of SRG technology, e.g., learning how to do three-body force evolution, and found that simplified models were invaluable in guiding the way to a fully realistic treatment and for providing generic insights~\cite{Jurgenson:2008jp}. 
In particular, one-dimensional treatments are feasible and preserve many of the important features (often because the linear algebra is basically the same). 
We follow this strategy here by extending the  one-dimensional treatment by Lipkin~\cite{lipkin1973quantum} of scattering,  first to an optical potential model with a two-level ``nucleus'' and then to a more characteristic (and finite range) potential. As in Ref.~\cite{Jurgenson:2008jp}, we use a one-dimensional interaction potential, originally introduced in Ref.~\cite{Alexandrou:1988jg}, that roughly simulates three-dimensional properties.

The paper is organized as follows.
Section~\ref{sec:formalism} lays out the background, formalism, and solution of the Lipkin one-dimensional scattering model~\cite{lipkin1973quantum}, the extension to its treatment with an optical potential in the Feshbach formalism, and the application of the SRG to this model. 
Results are given in Sect.~\ref{sec:results} for SRG decoupling and its consequences, the perturbativeness of the optical potential when evolved, and the manifestation of nonlocality. 
Section~\ref{sec:summary} summarizes our findings, puts them in a broader context, and points to future next steps.

\section{Formalism}\label{sec:formalism}
\subsection{One-dimensional model for scattering} \label{subsec:one-d_model}

Our theoretical laboratory for exploring SRG applied to optical potentials builds on the one-dimensional two-level model used by Lipkin~\cite{lipkin1973quantum} to demonstrate aspects of quantum scattering.%
\footnote{We use the conventions from Ref.~\cite{Landau:1996} rather than from Ref.~\cite{lipkin1973quantum}.}
In particular, the model has a particle scattering from a ``nucleus'' with a ground state and one excited state.
Thus we will have inelastic scattering with increasing energy of the projectile
and we can introduce an optical potential in the Feshbach formalism (see Sect.~\ref{subsec:opt_pot}).
We start with delta function interaction potentials as in Lipkin,  which enable analytic solutions to demonstrate the optical model and validate our numerical methods. We will generalize in Sec.~\ref{sec:results} to a nuclear-like potential as in Ref.~\cite{Jurgenson:2008jp} and work with numerical solutions.

First we review the solution from Ref.~\cite{lipkin1973quantum}.
The Hamiltonian for this scattering problem can be written in second quantization for the two-level system,
\begin{align}
    H &= \Kinetic -  \delta(x)
     [V_0 (a^\dagger_g a^{\pdag}_g + a^\dagger_e a_e^{\pdag}) 
     + V_1 (a^\dagger_g a^{\pdag}_e + a^\dagger_e a^{\pdag}_g)] \notag\\
   & \qquad \null + E_g a^\dagger_g a_g + E_e a^\dagger_e a_e ,
\end{align}
where $g$ and $e$ label the ground and excited states, respectively, and the kinetic energy is that of the scattered particle.
$V_0$ is the strength of the ground and excited state delta function potentials, $V_1$ is the strength of the coupling of the ground and excited state, and the energies of the ground and excited states are $E_g$ and $E_e$.
The coordinate-space Schr\"odinger equation for the projectile is
(henceforth in units with $\hbar^2/2m=1$):
\begin{align}
       \begin{pmatrix}
         \doublederivativeX + \kGround^2 & 0 \\
         0 & \doublederivativeX + \kExcited^2
       \end{pmatrix}
       &
       \begin{pmatrix}
        \vphantom{\doublederivativeX} \psiGround \\
        \vphantom{\doublederivativeX}  \psiExcited
       \end{pmatrix} 
       = 
       \notag \\
    &   -\delta(x)
       \begin{pmatrix}
        \vphantom{\doublederivativeX} V_0 & V_1\\ 
        \vphantom{\doublederivativeX} V_1 & V_0
       \end{pmatrix}
       \begin{pmatrix}
        \vphantom{\doublederivativeX} \psiGround \\
        \vphantom{\doublederivativeX} \psiExcited
       \end{pmatrix} ,
       \label{eq:matrixSeqn}
    \end{align}
where we have defined $\kGround$ and $\kExcited$ from:
\begin{align} 
      \kGround^2 &= E - E_g ,
      \label{eq:\kGround_def}
      \\
      \kExcited^2 &= E - E_e .
      \label{eq:\kExcited_def}
\end{align}

As we are considering a one-dimensional system, the solutions to \eqref{eq:matrixSeqn} can be classified as even or odd parity,. 
which is analogous to resolving the scattering wave function into partial waves in three dimensions.
For a delta function potential the odd solutions vanish at the origin, so the phase shifts are identically zero.
Hence we look for even-parity solutions for the ground state, 
\begin{equation}
    \psiGround = \alpha \cos\bigl(\kGround\abs{x} + \delta_0(E)\bigr) ,
    \label{eq:psi_wfn}
\end{equation}
and outgoing waves for $\psiExcited$:
\begin{equation}
 \psiExcited = \beta \exp(i\kExcited\abs{x}) .
\end{equation}
We can analytically solve \eqref{eq:matrixSeqn} to find~\cite{lipkin1973quantum} 
\begin{equation}
   \frac{\beta}{\alpha} =-\frac{V_1\cos(\delta_0)}{V_0 + 2i\kExcited} = \frac{2\kGround\sin(\delta_0)-V_0\cos(\delta_0)}{V_1} .
    \label{eq:alpha_beta}
\end{equation}
Solving \eqref{eq:alpha_beta} for the phase shift $\delta_0(E)$, we arrive at:
    \begin{equation}
        \tan(\delta_0) = \frac{V_0}{2\kGround} - \frac{V_1^2}{2\kGround(V_0 + 2i\kExcited)} .
        \label{eq:delta_phase}
    \end{equation}
For $E < E_e$, $k_e = i\sqrt{E_e - E}$.
    
\begin{figure*}[thb]
     \centering
     \subfloat[]{\includegraphics[width=0.9\columnwidth]{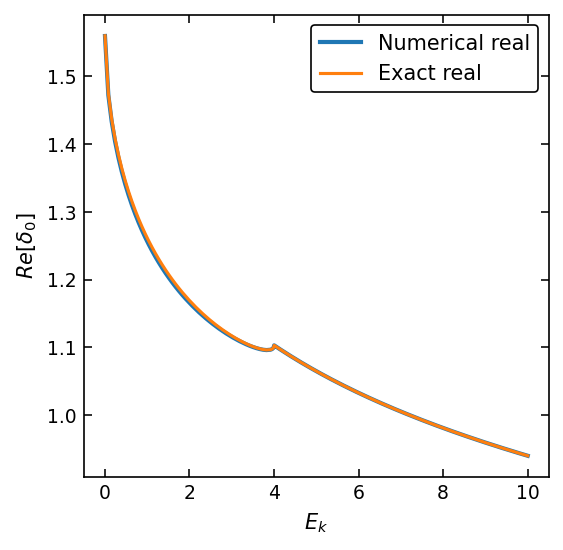}}%
     ~~~~~~
     \subfloat[]{\includegraphics[width=0.9\columnwidth]{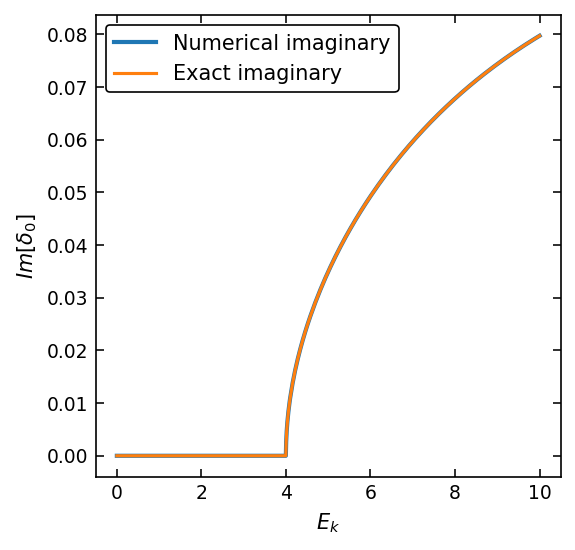}}
     \caption{(a) Real and (b) imaginary phase shifts of the two-level delta function potential with $E_g=0$,  $E_e=4$, $V_0 = 12$, and $V_1 = 7$ as a function of scattering energy $E_k = k^2$.
     The numerical solutions agree with the analytic result to high precision.}
     \label{fig:delta_fn_phase}
\end{figure*}

The phase shift is in general complex, as illustrated for the case of $E_g = 0$, $E_e = 4$, $V_0 = 12$, and $V_1 = 7$ in Fig.~\ref{fig:delta_fn_phase}.
The real phase shift is non-zero at zero energy because there are bound states and
at higher scattering energies slowly approaches zero.
The imaginary phase shift exhibits a clear inelastic threshold, i.e.\ the scattering energy where the projectile loses enough energy to excite the nuclear system to the higher energy level. 
Thus the threshold is given by $E_e - E_g$.

\subsection{Optical potential}\label{subsec:opt_pot}

We construct the one-dimensional optical potential for our model problem following the canonical Feshbach formalism~\cite{Feshbach:1958nx,Feshbach:1958wf,Feshbach:1962ut}.
We introduce idempotent projection operators $P$ and $Q$, where $P$ projects onto the ground state of the potential and $Q$ projects onto the excited state.
The optical potential then takes the same form as in three dimensions:
\begin{equation}            \label{eq:feshbach_opt_pot}
    \Vopt(E) = PVP + PVQ\frac{1}{E-QHQ+i\epsilon}QVP .
\end{equation}
Starting from a two-level potential $V$, the first term describes elastic scattering (i.e., the system stays in the ground state) while the second term accounts for scattering that excites the nucleus into the second level. 
In the following, we will usually suppress the explicit energy argument for $\Vopt$. 

In detail for the delta function case, the full potential in \eqref{eq:feshbach_opt_pot} is 
\begin{equation}
    V(x) = -\delta(x)\begin{pmatrix}
           V_0 & V_1\\ 
           V_1 & V_0 ,
          \end{pmatrix} ,
\end{equation}
so we identify $PVP= -V_0\delta(x)$ and $PVQ= QVP = -V_1\delta(x)$.
To solve for $\Vopt(x)$, we introduce $\Gtilde(x)$ as the solution to
\begin{equation}
      \bigl(E - E_e  + \doublederivativeX + 
          V_0\delta(x) + i\epsilon\bigr) \Gtilde(x)  = V_1\delta(x) 
          \label{eq:Green's_fn_lipkin}
\end{equation}
with outgoing wave boundary conditions for $\Gtilde(x)$ encoded by the $+i\epsilon$ as in \eqref{eq:feshbach_opt_pot}, so that
\begin{equation} \label{eq:Vopt_Gtilde}
    \Vopt(x) = -V_0\delta(x) + V_1\delta(x)\Gtilde(0) .
\end{equation}
Integrating \eqref{eq:Green's_fn_lipkin} across the delta function and matching to plane wave solutions for $x\neq 0$, we find
\begin{equation}
    \Gtilde(x) = 
      \begin{cases}
        e^{i\kExcited x} + \Bigl(\frac{\ts V_1}{\ts V_0 + 2i\kExcited} - 1 \Bigr)
        e^{-i\kExcited x} & x\leq 0 , \\[10pt]
        \frac{\ts V_1}{\ts V_0 + 2i\kExcited}e^{i\kExcited x} & x\geq 0 ,
      \end{cases} 
      \label{eq:G_fn_regions}
\end{equation}
so that the final analytic form for the optical potential is
\begin{equation}
  \Vopt(x) = \left( -V_0 + \frac{V_1^2}{V_0 + 2i\kExcited} \right) \delta(x) . 
\end{equation}

To find the phase shifts, we return to the Schr\"odinger equation and substitute $\Vopt$:
\begin{equation}
  -\dv[2]{\Psi}{x} + \left( -V_0 + \frac{V_1^2}{V_0 + 2i\kExcited} \right) \delta(x)\Psi(x) = \kGround^2\Psi(x) .
\end{equation}
As already noted, odd parity states vanish at the origin so the phase shift for such states are zero. 
Considering  only even parity states, we solve the Schr\"odinger equation to find the wavefunction $\Psi(x)$ using \eqref{eq:psi_wfn}.
After matching at $x=0$,
we arrive at the phase shift for the even parity states:
\begin{equation} \label{eq:phase_shift_delta_pot}
  \tan(\delta_0) = \frac{V_0}{2\kGround} - 
  \frac{V_1^2}{2\kGround(V_0 + 2i\kExcited)} ,
\end{equation}
which, of course, is the same as the original phase shift expression derived for the full two-level potential in \eqref{eq:delta_phase}.

Alternatively, we can work in momentum space.
We define the Green's functions
\begin{align}
    \Gzerog(k,k';E) &= \frac{\delta(k-k')}{\kGround^2 - k^2 +i\epsilon} , \label{eq:Gg} \\
    \Gzeroe(k,k';E) &= \frac{\delta(k-k')}{\kExcited^2 - k^2 +i\epsilon} ,
    \label{eq_Ge}
\end{align}
and insert complete sets of momentum states in \eqref{eq:feshbach_opt_pot}:
\begin{equation}
     \matrixel{p}{\Vopt}{p'} = -\frac{V_0}{2\pi} + \left(\frac{V_1}{2\pi}\right)^2 \int dk \int 
    dk' \matrixel{k}{G_e}{k'} .
\end{equation}
We relate $G_e$ to the free-space Green's function $\Gzeroe$:
\begin{align}
  G_e &= \Gzeroe - \Gzeroe V_0 \Gzeroe + \Gzeroe V_0 \Gzeroe V_0 \Gzeroe
  + \dots \notag  \\
    &= \Gzeroe - \Gzeroe V_0 G_e .
  \label{eq:Greens_function_eq}
\end{align}
We can sum the expansion of the energy-dependent term:
\begin{align}
  \int\! dk \! \int\! dk' \matrixel{k}{G_e}{k'} &= \int dk 
  \frac{1}{E - E_e - k^2+ i\epsilon}\notag\\ &\hspace*{-20pt}\null- \frac{V_0}{2\pi} \left( \int 
  dk\frac{1}{E - E_e - k^2+ i\epsilon} \right)^2 + 
  \dots \notag \\
  &= \frac{2\pi}{V_0 + 2i\kExcited} .
\end{align}
Thus we obtain the same optical potential as before, but now in momentum space:
\begin{equation} \label{eq:Vopt_delta_momspace}
     \matrixel{k}{\Vopt(E)}{k'} = \frac{1}{2\pi}\left(-V_0 + \frac{V_1^2}{V_0 + 
     2i\kExcited}\right) .
\end{equation}

To extract the phase shifts, we use an operator Lippman-Schwinger equation for the T-matrix, %
\begin{equation}
    T = \Vopt + \Vopt \Gzerog T ,
\end{equation}
which can be rearranged to yield:
\begin{align} \label{eq:T_matrix_delta}
    \matrixel{p}{T}{p'} &=  \matrixel{p}{\Vopt}{p'}
    \notag\\ \null &
 \quad \times \frac{1}{1 - \frac{2\pi}{2i\kGround} 
  \matrixel{p}{\Vopt}{p'}} .
\end{align}
Using the relation $2\pi\!\matrixel{p}{T}{p} = -2 \kGround e^{i\delta_0} \sin(\delta_0)$, we arrive again at the phase shift expression \eqref{eq:phase_shift_delta_pot}.
The optical potential construction will have to be performed numerically when we switch to a more realistic finite-range or once we do any SRG evolution; this is detailed in Sec.~\ref{subsec:SRG_opt_calc}.

\subsection{SRG evolution}\label{subsec:SRG}

The SRG flow equation,
\begin{equation}
    \frac{dH(s)}{ds} = \comm{\eta(s)}{H(s)} ,
    \label{eq:flow_eq}
\end{equation}
is an operator equation that implements a continuous series of unitary transformations, such that operators are modified while keeping observables invariant \cite{Furnstahl:2013oba}.
The SRG evolution is parameterized by the 
flow parameter $s$ or $\lambda = s^{-1/4}$, where
$\lambda$ can be viewed as setting the resolution scale of the physical problem at hand.
Note that Eq.~\eqref{eq:flow_eq} can be evaluated in any basis and has the same form in one or three dimensions.
In one-dimension, we evolve separately the even and odd components of the potential (see Sec.~\ref{subsec:nonlocality}).

The pattern of the RG flow is dictated by the generator $\eta$.
For example, the Wegner generator is $\eta(\lambda) = \comm{H_d(\lambda)}{H_{od}(\lambda)}$, i.e., it is the commutator of the diagonal ($d$) and off-diagonal ($od$) components of the Hamiltonian $H$.
In momentum space, evolution with this generator decouples high- and low-relative momenta in the Hamiltonian, hence shifting from a high-RG resolution picture to a low-RG resolution one. 
At high resolution, convergence of basis expansions of the wave function is significantly slowed down because of high-momentum components in low-energy wavefunctions (referred to generically as short-range correlations or SRCs).  
Evolving to low-resolution significantly softens the Hamiltonian, making it more perturbative, which enables simpler calculations of low-energy observables. Nevertheless, a consequence of momentum decoupling is that the SRG flow introduces nonlocality to the interactions; in Sec.~\ref{subsec:nonlocality} we will contrast this with the nonlocality of the optical potential.

The Wegner generator drives the Hamiltonian to band-diagonal form.
A band-diagonal SRG evolution \emph{locally} decouples momentum~\cite{Dainton:2013axa}, where the width of the diagonal is plotted versus momentum squared is roughly $\lambda^2$.
An alternate decoupling scheme is block decoupling, where the momentum matrix elements of the potential only coupled if the momenta are on the same side of a specified cutoff momentum $\Lambda$~\cite{Anderson:2008mu}.
A block decoupled potential is achieved using the SRG generator:
\begin{align}
    \eta(\lambda) = \comm{P_\Lambda H(\lambda)P_\Lambda + Q_\Lambda H(\lambda)Q_\Lambda}{H(\lambda)} .
\end{align}
Here the projection operators $Q_\Lambda$ and $P_\Lambda$ partition the potential above and below the cutoff $\Lambda$~\cite{Hergert:2016iju}. 
(Note: these are unrelated to the $P$ and $Q$ projection operators from the optical potential.) 

\subsection{Calculating Optical Potential at Each Resolution Scale}\label{subsec:SRG_opt_calc}

We construct the optical potential at each SRG resolution scale by revisiting the Feshbach formalism for the optical potential \eqref{eq:feshbach_opt_pot}:
    \begin{align}\label{eq:feshbach_opt_pot2}
        \Vopt = V_{00} + V_{01}G_{11}V_{10} ,
    \end{align}
where the terms:
    \begin{align}
        V_{00} &= PVP , \notag\\
        V_{01} &= V_{10} = PVQ,
        \label{eq:Feschbach with sub}
    \end{align}
are extracted from the two-level potential introduced in \ref{subsec:one-d_model}. The potentials $V_{00}$,$V_{01}$, and $V_{10}$ are SRG-evolved components of the original potential. We identify $G_{11}$ from \eqref{eq:feshbach_opt_pot} as $G_{11} = Q(E-H+i\epsilon)^{-1}Q$.

To treat the integrations over the Green's function numerically in momentum space, we first
expand $G_{11}$ in terms of the free-space Green's function $G^0_{11}$ from \eqref{eq_Ge}:
    \begin{align}
        G_{11} = G^0_{11} + G^0_{11}V_{11}G^0_{11} + G^0_{11}V_{11}G^0_{11}V_{11}G^0_{11} + ... 
        \label{eq:G_11 expansion}
    \end{align}
where we suppress momentum dependence and intermediate integrations. Substituting \eqref{eq:G_11 expansion} into \eqref{eq:feshbach_opt_pot2} we can write the optical potential as:
\begin{align} \label{eq:Vopt_with_M}
        \Vopt(E) = V_{00} + V_{01} G^0_{11}(E) \bigl(V_{10} + \wt M(E)\bigr) .
\end{align}
 where 
 \begin{align}
  \label{eq:Mtilde_eq}
    \wt M(E) = \wt M_0(E) + V_{11} G^0_{11}(E) \wt M(E) , 
 \end{align} and $M_0(E) \equiv V_{11} G^0_{11}(E) V_{10}$.
We now look at two regimes for $G^0_{11}$ in momentum space.
For negative $\kExcited^2$, $G^0_{11}$ does not have a singularity.
For positive $\kExcited^2$, we split the Green's function into a principal value and on-shell imaginary part:
    \begin{align}
        \langle p' | G^0_{11} | p \rangle = \left[\frac{\mathcal{P}}{\kExcited^2 - p^2} - i\pi\delta(\kExcited^2-p^2)\right] \delta(p-p') .
    \end{align}
We solve numerically by discretizing $\wt M(E)$ using a Gaussian quadrature mesh with weights $w_i$, 
    \begin{align}
        (\wt M)_{ij} = (\wt M_0)_{ij} + 4\sum_{l,m=1}^{N} w_l  w_m
        (V_{11})_{il} (G^0_{11})_{lm} (\wt M)_{mj} .
    \end{align}
We calculate the real and imaginary components of $\wt M_0$ by looking at the two cases of $\kExcited^2$:
    \begin{align}
        \text{Re}(\wt M_0)_{ij} =
        \begin{cases}
         2\sum_{l=1}^{N} w_l (V_{11})_{il} \ds\frac{1}{\kExcited^2 - k_l^2} (V_{10})_{lj} & \kExcited^2 < 0 \\
         2\sum_{l=1}^{N} w_l \ds\frac{1}{\kExcited^2 - k_l^2} 
         \bigl(
            (V_{11})_{il}(V_{10})_{lj}\\ \qquad \null- (V_{11})_{iN+1}(V_{10})_{N+1j}
         \bigr) & \kExcited^2 > 0
         \end{cases}
    \end{align}
and 
    \begin{align}
        \text{Im}(\wt M_0)_{ij} =
        \begin{cases}
          0 & \kExcited^2 < 0 \\
          -\ds\frac{\pi}{\kExcited} (V_{11})_{iN+1}(V_{10})_{N+1j} & \kExcited^2 > 0
        \end{cases}
    \end{align} 
We use these discretized forms of $\wt M$ and $\wt M_0$ to construct $\Vopt(E)$.

\section{Results}\label{sec:results}

The general goal of the present investigations is to examine to what extent the characteristic 
features of SRG evolution of potentials are inherited by the corresponding optical potential.
In the next three sections, we examine in turn decoupling in the potential, the consequences of that decoupling on perturbativeness, and finally the nature of nonlocality inherent in the optical potential and that induced by the SRG.
We show results in the main document for the even parity states only; some corresponding results for the odd parity states are given in the Supplemental Material (see the Appendix).

For the bulk of our explorations we use an interaction that is the sum of an attractive and a repulsive Gaussian~\cite{Jurgenson:2008jp},
which we will call the ``two-Gaussian potential''. 
In momentum space this is:
\begin{equation}
            V(k,k') = \frac{V_1}{2\pi}e^{-(k-k')^2\sigma_1^2} + \frac{V_2}{2\pi}e^{-(k-k')^2\sigma_2^2},
\end{equation}
with parameter values given in Table~\ref{tab:Negele_parameters};
the ``Diagonal'' values were chosen in Ref.~\cite{Alexandrou:1988jg} to reflect empirical three-dimensional nuclear properties in this one-dimensional model.
We use ``Off-Diagonal I" values everywhere except for Figs.~\ref{fig:Weinberg_2lvl_repulsive} and~\ref{fig:Weinberg_2lvl_opt_pot_repulsive}, for which we use the values for ``Off-Diagonal II".
Our analysis is not sensitive to the particular values of these parameters.
\begin{table}[htb]
	\caption{%
	Parameters for the two-Gaussian potential. For the one-level system, we use only the parameters of the ground-state potential.
	}
	\label{tab:Negele_parameters}
	\begin{ruledtabular}
		\begin{tabular}{c|cccc}
		   Potential Level & $V_1$ &  $V_2$ &  $\sigma_1$ & $\sigma_2$ \bigstrut\\
		    \colrule
      			Diagonal & $12$ &  $-12$ & $0.2$ & $0.8$ \null \bigstrut\\
      			Off-Diagonal--I & $7.5$ & $-7.5$ & $0.2$ & $0.8$ \\
  		  Off-Diagonal--II & $2.0$ & $-2.0$ & $0.2$ & $0.8$ \\
		\end{tabular}
  	\end{ruledtabular}
\end{table}

\subsection{SRG decoupling analysis and consequences}\label{subsec:decoupling}

\begin{figure*}[tbh]
     \subfloat[]{\includegraphics[width=0.46\textwidth]{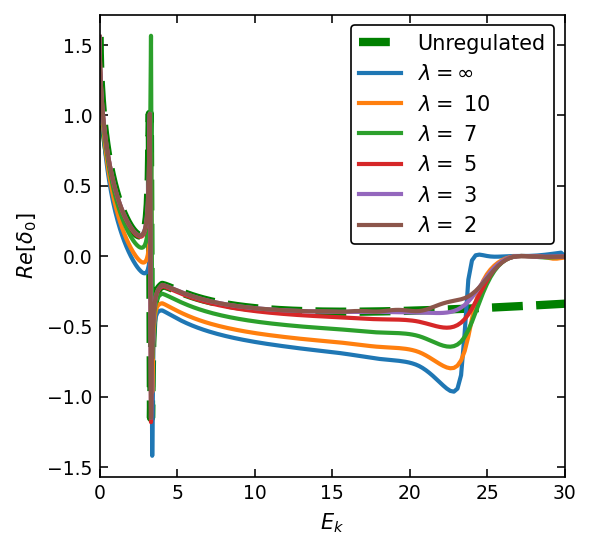}}
    \hspace{0.5cm}
    \subfloat[]{\includegraphics[width=0.46\textwidth]{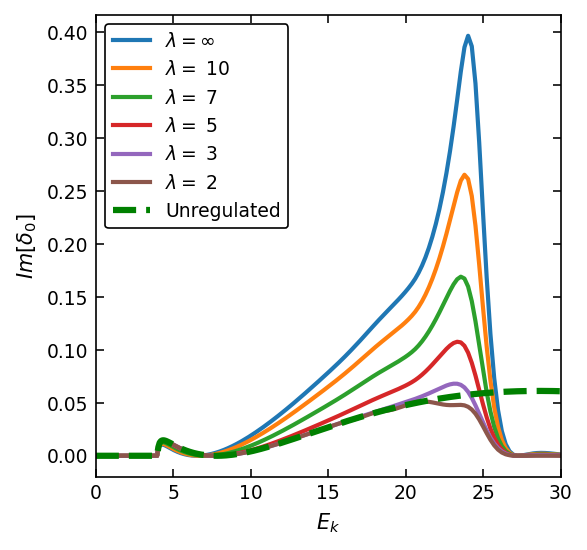}}
    \caption{Test of decoupling using (a) real and (b) imaginary even-parity phase shifts of the optical potential under SRG evolution, as constructed from the two-level two-Gaussian potential (see Table~\ref{tab:Negele_parameters}), with $E_g = 0$ and $E_e = 4$. 
    The SRG-evolved potential for each curve labeled by an SRG $\lambda$ value is cut off by a regulator at $k=\sqrt{25}$ (corresponding to $E_k = 25$).
    The unevolved and undistorted phase shifts are labeled ``unregulated''.
    For $\lambda$ above the cutoff, the phase shifts are modified, but converge rapidly to the unevolved phase shifts as the SRG resolution $\lambda$ is decreased below the cutoff. (Note: without a regulator, all curves are identical for all $E_k$.)} 
    \label{fig:Regulator_1}
\end{figure*}

\begin{figure*}[tbh]
    \subfloat[]{\includegraphics[width=0.46\textwidth]{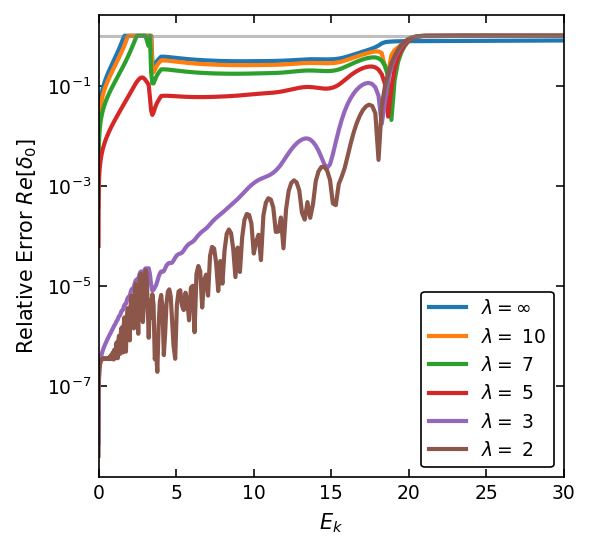}}
    \hspace{0.5cm}
    \subfloat[]{\includegraphics[width=0.46\textwidth]{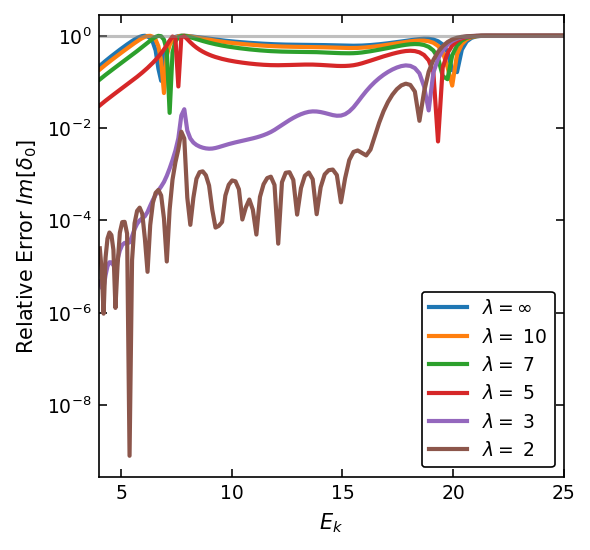}}
    \caption{Same as Fig.~\ref{fig:Regulator_1} but plotting the relative errors of the even-parity phase shifts from regulated SRG-evolved optical potentials compared to the exact phase shifts.
    The light gray horizontal line marks 100\% relative error.} 
    \label{fig:Regulator_2}
\end{figure*}

\begin{figure*}[tbh]
    \subfloat[]{\includegraphics[width=0.33\textwidth]{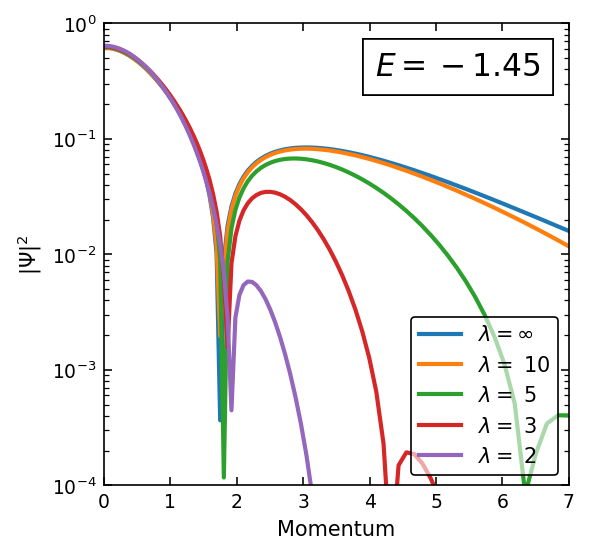}}%
    \subfloat[]{\includegraphics[width=0.33\textwidth]{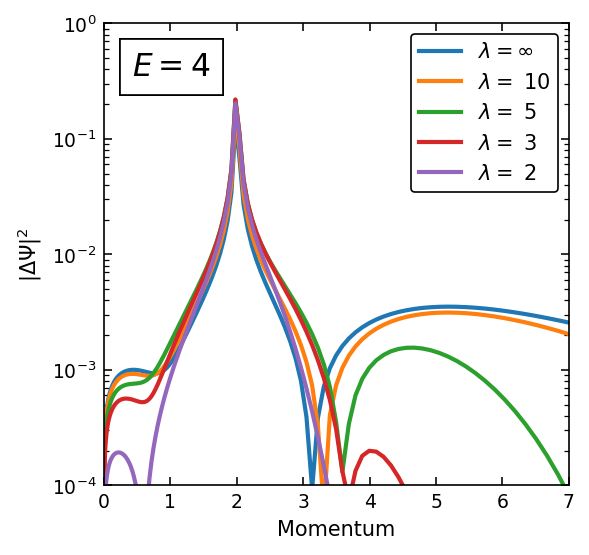}}%
    \subfloat[]{\includegraphics[width=0.33\textwidth]{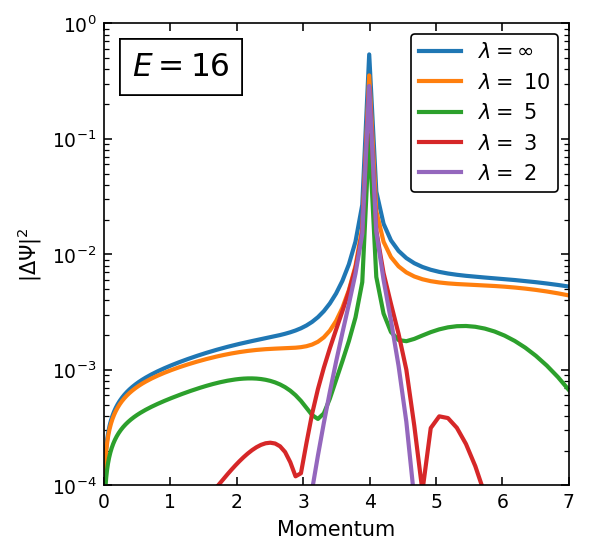}}
    \caption{Optical potential wavefunctions evaluated at energies (a) $E = -1.44$, (b) E = 4, and (c) E = 16 for the two level two-gaussian potential with levels  $E_g = 0$ and $E_e = 4$. We SRG evolve the two-Gaussian potential to  several $\lambda$ values from which we construct the optical potential. Local decoupling of the SRG is exhibited by the suppression of momentum modes of the scattering wavefunctions above and below the scattering momentum $\sqrt{E}$. } 
    \label{fig:wf_1}
\end{figure*}

As described in Sec.~\ref{subsec:SRG}, the Wegner generator drives the Hamiltonian to band-diagonal form.
We will first look at the consequences of this RG evolution pattern for phase shifts calculated with the optical potential of the two-Gaussian potential.
We calculate the phase shifts by numerically solving the Lippman-Schwinger equation for the T-matrix in one-dimension~\cite{Landau:1996}.

From early work on the SRG with (real) nucleon-nucleon potentials~\cite{Jurgenson:2007td}, we know that phase shifts at energies corresponding to relative momenta below the SRG $\lambda$ only get substantial contributions from matrix elements of the potential with arguments less than  $\lambda$. 
This effect of decoupling was manifested in Ref.~\cite{Jurgenson:2007td} by introducing an artificial regulator that smoothly cut off the potential at a specified cutoff, so that matrix elements with at least one argument above the cutoff were set to zero. 
The signature of decoupling was that the phase shift was equal to the original phase shift for energies below the imposed cutoff and zero above. 
As expected, the same is seen to be true with our one-dimensional potentials when SRG-evolved.

A representative example of the corresponding test of decoupling for the optical potential in our one-dimensional model is shown in Figs.~\ref{fig:Regulator_1} and \ref{fig:Regulator_2}.
The curves labeled ``Unregulated'' are the (a) real and (b) imaginary even-parity phase shifts calculated from the original unevolved two-Gaussian potential (see Table~\ref{tab:Negele_parameters}), with no regulator applied.
The zero-energy phase shift of ${\pi}/{2}$ is what is expected from Levinson's theorem for even-parity one-dimensional systems~\cite{Barton_1985}, as there is one bound state for this choice of potential with $E_g =0$ and $E_e = 4$.
The expected inelastic threshold of $E_k = 4$ is clear in the right panel and also shows up as a kink in the left panel.

The other curves in Fig.~\ref{fig:Regulator_1} are the phase shifts of the \emph{regulated} optical potentials constructed from the two-level two-Gaussian potential at the indicated SRG $\lambda$.
We emphasize that, if unregulated, all of the phase shifts from the optical potentials at every SRG resolution are the same, because the SRG transformations are unitary.
However, with a regulator at $E_k = 25$ (thus a momentum cutoff of $k=5$) we see that the $\lambda = \infty$ (i.e., unevolved) phase shifts, both real and imaginary, disagree significantly from the ``unregulated'' phase shifts over the full range of energies plotted.
This demonstrates the important role of momentum matrix elements above the regulator scale for the unevolved potential, even for the lowest energies well below that scale.
But with SRG evolution, the phase shifts become less distorted and agree closely with the unregulated observables once $\lambda$ is less than the regulator scale.
To more clearly see the decoupling effects of the SRG evolution on the phase shifts, we shift to a log-log plot for the real phaseshifts and the corresponding relative phase shift error in Fig.~\ref{fig:Regulator_2}. 
Thus decoupling implies that once $\lambda^2$ is below the cutoff energy (so here for $\lambda = 3$ and $\lambda = 2$), those omitted high-momentum matrix elements become irrelevant for optical potential phase shifts, just as with any nucleon-nucleon potential.

The consequences of decoupling as a function of resolution scale should also be manifested in the bound state and scattering wave functions. 
In Ref.~\cite{Jurgenson:2008jp}, suppression of high-momentum tails of the bound-state wave function (i.e., SRCs) for the two-Gaussian potential was observed as the potential was SRG-evolved to lower resolution scales.
In Fig.~\ref{fig:wf_1}(a) we see this same signature of decoupling for the bound state wavefunction from the optical potential derived from the two-level two-Gaussian potential.
The logarithm of the wavefunction squared shows the same pattern of high-momentum suppression as seen in momentum distributions of nuclei~\cite{Tropiano:2021qgf}, with the start of suppression just above the value of SRG $\lambda$.
In panels (b) and (c) are plotted the logarithm of the scattering wavefunction squared for two energies, where the on-shell delta function is omitted to define $\Delta\psi$~\cite{Landau:1996}.
We again see high-momentum tails getting suppressed as we evolve to lower $\lambda$ values.
But in addition the \emph{lower} momentum components of the wavefunction also get suppressed due to the effect of \emph{local} decoupling from the band-diagonal SRG evolution~\cite{Dainton:2013axa,More:2017syr}.

In summary, the SRG decoupling behavior of the optical potential, observables, and wave functions is completely aligned with what is observed for free-space nucleon-nucleon potentials.
Therefore we expect the consequences described in Refs.~\cite{More:2015tpa,More:2017syr,Tropiano:2021qgf,Tropiano:2022jjj} to follow as well, as summarized in Sec.~\ref{sec:summary}.

\subsection{Perturbativeness of the optical potential}\label{subsec:perturbativeness}

\begin{figure}[tbh]
    \includegraphics[scale=.75]{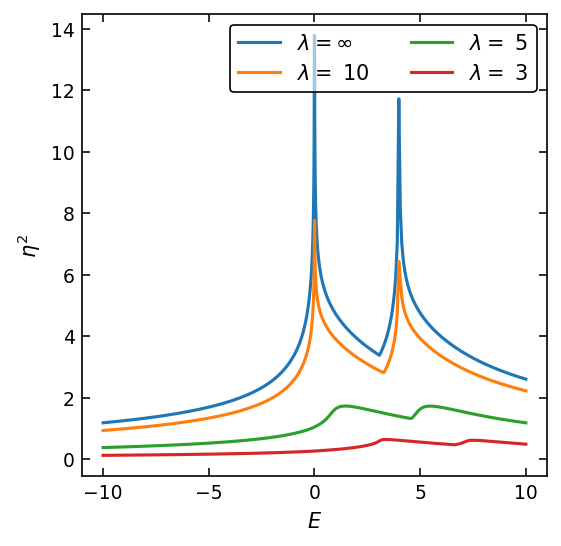}
        \caption{Repulsive Weinberg eigenvalues for the two-level two-Gaussian potential using ``Off-diagonal--II'' values (see Table~\ref{tab:Negele_parameters}). We plot the absolute value squared $|\eta|^2$ of the complex eigenvalue versus the energy for a range of SRG $\lambda$ resolutions. The initial peaks are at the values of $E_g=0$ and $E_e=4$.} 
        \label{fig:Weinberg_2lvl_repulsive}
\end{figure}

\begin{figure}[tbh]
    \includegraphics[scale=.75]{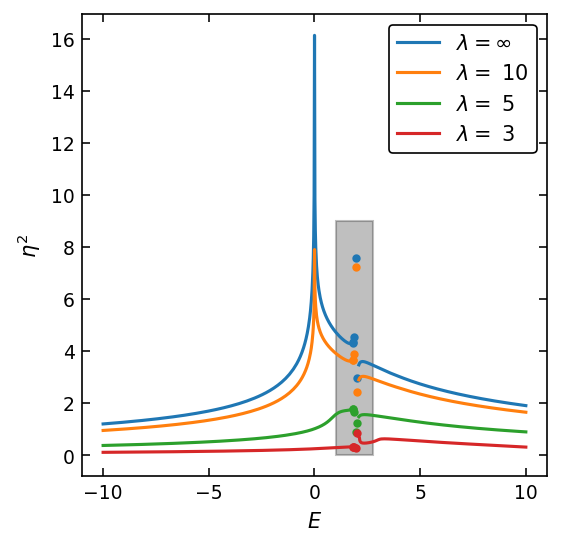}
    \caption{Same as Fig.~\ref{fig:Weinberg_2lvl_repulsive} but for the repulsive eigenvalues derived from  the  optical potential at each $\lambda$. We exclude from consideration the narrow singular region for Weinberg eigenvalues at energies in the gray boxed region, which are problematic for classifying as repulsive or attractive.} 
        \label{fig:Weinberg_2lvl_opt_pot_repulsive}
\end{figure}

At high RG-resolution in three dimensions, such as when using the Argonne V18 potential~\cite{Wiringa:1994wb}, a strong short-range repulsive potential and the short-range tensor force render the nuclear many-body problem highly nonperturbative~\cite{Hergert:2016iju,Furnstahl:2013oba,Bogner:2006tw}.
For scattering, this is manifested as contributions to the Born series in the lower partial waves getting significantly larger with each successive order.
We can identify the source of this nonperturbativeness as the mixing of high-momentum components into low-energy states driven by the short-range interactions.
As SRG evolution decouples this mixing, we expect and observe that low resolution Hamiltonians are more perturbative, meaning that successive terms in the Born series and many-body perturbation theory (MBPT) do not grow as fast.
Note that nonperturbative behavior from the existence of bound states (e.g., the deuteron or pairing), are not changed by unitary SRG evolution 
(when combined with Pauli blocking in nuclei, MBPT expansions do become largely perturbative~\cite{Bogner:2005sn}).

From the decoupling with SRG evolution of the optical potential verified in the last section, we expect this potential will also inherit the more perturbative trend with lower resolution.
To manifest and quantify this perturbativeness, we adapt to our one-dimensional problem the method of ``Weinberg eigenvalues,'' which are the eigenvalues of the $G^0 V$ matrix that appears in the Lippmann-Schwinger Born series for scattering.
Among these eigenvalues are ones associated with the repulsive parts of the potential and ones associated with the attractive parts of the potential; we focus on the former.
The treatment of Weinberg eigenvalues in one dimension largely carries over from three dimensions, with the exception of additional singular behavior due to the different measure in momentum space.

\begin{figure*}[tbh]
    \subfloat[]{\includegraphics[width=0.33\textwidth]{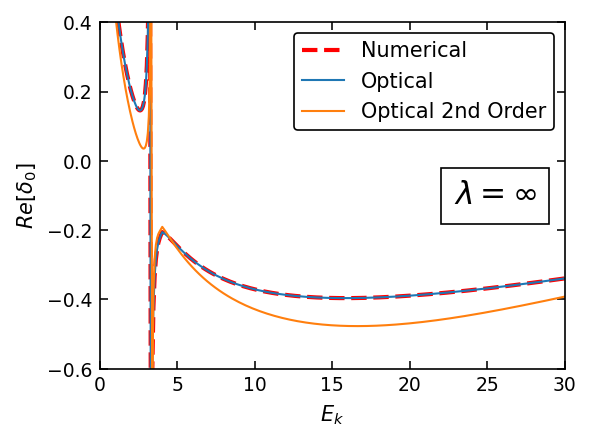}}%
    \subfloat[]{\includegraphics[width=0.33\textwidth]{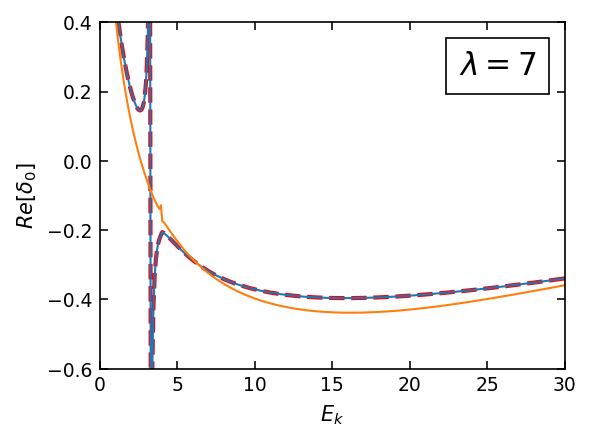}}%
    \subfloat[]{\includegraphics[width=0.33\textwidth]{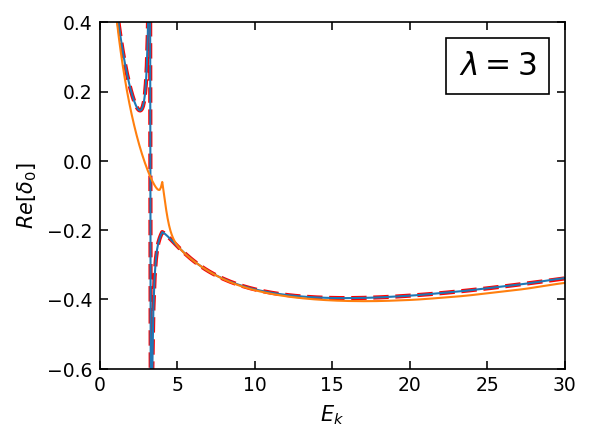}}\\
    \subfloat[]{\includegraphics[width=0.33\textwidth]{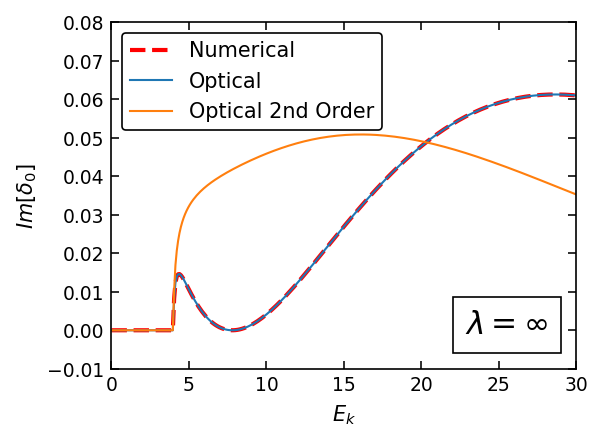}}%
    \subfloat[]{\includegraphics[width=0.33\textwidth]{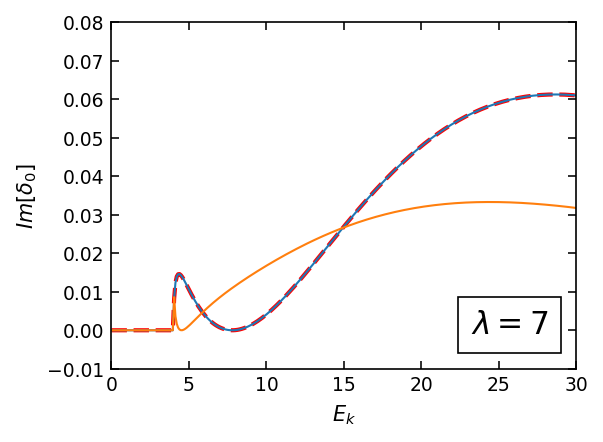}}%
    \subfloat[]{\includegraphics[width=0.33\textwidth]{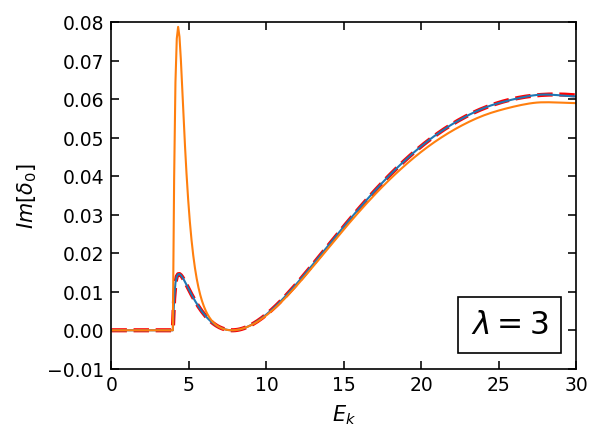}}
    \caption{Phase shifts two-level two-Gaussian system calculated from the full optical potential (``Optical'') compared to the direct calculation and to the optical potential calculated to second order (``Optical 2nd Order''). The real phase shifts are in panels (a), (b), and (c) for SRG $\lambda = \infty$, 7, and 3, and the imaginary phase shifts are in panels (d), (e), and (f) for the same $\lambda$'s.} 
    \label{fig:phases_opt_truncated}
\end{figure*}

The Weinberg eigenvalues $\eta(z)$ at complex energy $z$ are defined by~\cite{Weinberg:1963zza,Ramanan:2007bb,Hoppe:2017lok}
\begin{equation}
    G^0(z)V\ket{\Psi_\eta} = \eta(z) \ket{\Psi_\eta} ,
    \label{eq:Wein_eig_eq}
\end{equation}
but in practice we will want to integrate over the singular Green's function, so we can solve for the left eigenvalues of $G^0 V$ or the right eigenvalues of
\begin{equation} \label{eq:Wein_eig_eqp}
    V G^0(z) \ket{V\Psi_\eta} = \eta(z) \ket{V\Psi_\eta} ,
\end{equation}
which has the same spectrum.
If we apply expansion of the T-operator at $z$:
\begin{equation}
    T(z) = V + VG^0(z)V + VG^0(z)VG^0(z)V + \ldots
\end{equation}
to $\ket{\Psi}$ and use \eqref{eq:Wein_eig_eq},
we obtain:
\begin{equation} \label{eq:Wein_eig_eq2}
    T(z)\ket{\Psi} = V (1 + \eta(z) + \eta(z)^2 + \eta(z)^3 + \ldots)\ket{\Psi} .
        \end{equation}
For this ordinary power series in $\eta(z)$ to converge, $\eta(z) < 1$. 
If this is the case for all $\eta(z)$, the interaction $V$ is perturbative at this energy, and the rate of convergence is determined by the largest such $\eta(z)$.
These discrete eigenvalues are defined in the complex $z$ plane with a cut along the positive real axis.

Equation~\eqref{eq:Wein_eig_eq} can be rearranged to take the form of the Schr\"odinger equation with a modified potential:
\begin{equation}
    \Bigl(H_0 + \frac{V}{\eta(z)}\Bigr)\ket{\Psi}= z\ket{\Psi} .
\end{equation}
This shows that $\eta(z)^{-1}$ acts like a complex, energy-dependent coupling constant that multiplies the interaction $V$.
If there is a bound state at $z = E_b$, then $\eta(E_b) = 1$.
For real $z = E < 0$, a purely attractive $V$ has only positive eigenvalues, while a purely repulsive $V$ has only negative eigenvalues.
For a mixed attractive and repulsive potential, we denote the eigenvalues as attractive or repulsive according to the sign.
For positive energies $E$ we take $z = E+i\epsilon$ and the modified Schr\"odinger equation has complex eigenvalues.
They are classified as attractive or repulsive depending on the continuation of $\eta(z)$ from $z$ on the negative real axis.

To calculate the Weinberg eigenvalues for the two-level system, we solve \eqref{eq:Wein_eig_eq} or \eqref{eq:Wein_eig_eqp} with $z = E+i\epsilon$, accounting for the $2\times2$ matrix structure in momentum representation:
\begin{align}
    \int\! dk'
    &\begin{bmatrix}
       V_{00}(k,k') \Gzerog(E) & V_{01}(k,k') \Gzerog(E)\bigstrut
       \\
       V_{10}(k,k') \Gzeroe(E) & V_{11}(k,k') \Gzeroe(E)\bigstrut 
    \end{bmatrix}\braket{k'}{V\Psi_\eta(z)} 
    \notag \\
    &= \eta(z)\braket{k}{V\Psi_\eta(z)} ,
    \label{eq:Weinberg_matrix_equation}
\end{align}
where $\braket{k}{V\Psi_\eta(z)}$ is a two-component vector wavefunction and we have suppressed the momentum indices of the
Green's functions $\Gzerog(k,k';E)$ and $\Gzeroe(k,k';E)$ from Eqs.\eqref{eq:Gg} and \eqref{eq_Ge}.
For the optical potential we have only the 00 component with the energy-dependent potential in place of $V_{00}$.

The repulsive Weinberg eigenvalues for the full two-level two-Gaussian potential are shown for a range of energies at several SRG $\lambda$ values in Fig.~\ref{fig:Weinberg_2lvl_repulsive}.
The appearance of the two initial ($\lambda = \infty$) peaks at energies $E_g$ and $E_e$ is unfamiliar from three-dimensional investigations.
The source of these peaks are singularities from the Green's functions in Eq.~\eqref{eq:Weinberg_matrix_equation}, which
in three dimensions are suppressed by the measure in momentum integrals.
Discounting this feature, the general pattern we are looking for, namely a dramatic suppression of the repulsive eigenvalues with decreasing $\lambda$, is clear.
In Fig.~\ref{fig:Weinberg_2lvl_opt_pot_repulsive} we see that the same pattern is seen for the optical potential.

A key question now is whether the increased pertubativeness at lower RG resolutions is reflected in improved convergence of Eq.~\eqref{eq:Vopt_with_M}. 
To study this we compare in Fig.~\ref{fig:phases_opt_truncated} the phase shifts from the full optical potential (``Optical'') to those obtained with \eqref{eq:Vopt_delta_momspace} truncated at the $V_{01}G^0_{11}(E)V_{10}$ term (``Optical 2nd Order'').
For the unevolved optical potential ($\lambda = \infty$), there are substantial differences, particularly for the imaginary phase shifts.
With evolution to $\lambda = 7$ (panel (b)), the agreement for both real and imaginary parts is significantly improved while the results for $\lambda = 3$ in panel (c) show excellent convergence except near the inelastic threshold ($\lambda = 2$ is even better, although more prone to numerical artifacts).

\subsection{Nonlocality} \label{subsec:nonlocality}

The formal structure of an optical potential as in Eq.~\eqref{eq:feshbach_opt_pot} implies a spatial nonlocality, but because this is a numerical complication, phenomenological optical potentials have typically assumed a local energy-dependent form~\cite{Dickhoff:2018wdd}.
However,
several recent studies have reexamined the importance of explicit nonlocality 
when calculating reaction observables~\cite{Titus:2016qmt,Li:2018zdo}. 
In comparison to a local optical potential, it was found that introducing nonlocality drastically improved the accuracy of experimental $(d,p)$ transfer cross sections, specifically the calculation of spectroscopic factors.
The latter are scale and scheme dependent quantities, so it is important to understand the impact of nonlocality.
At the same time, it was also found that some observables, such as the spin distribution for the nonelastic transfer cross section for $(d,p)$ processes, are largely unaffected.
Clearly there is more to be understood from the purely phenomenological perspective.

Here we seek insight into the nature of optical potential nonlocality in the RG framework by examining our model system at different resolutions.
This will enable us to contrast the nonlocality arising in the Feshbach formalism with that induced from SRG evolution.
We start by examining the nonlocality purely from the SRG by evolving a completely local potential, the two-Gaussian potential (i.e., treating this potential as the interaction potential between two nucleons or as the potential for purely elastic scattering). 
The nonlocality can be manifested in one dimension by
considering the dependence on $p$ and $q$ defined as:
\begin{align}
  p = \frac{k + k'}{2} ,
        \\
 q = k - k' .
\end{align}
In this representation, $q$ is the momentum transfer and $p$ serves as a ``nonlocality parameter,'' which will be used to directly quantify the degree of nonlocality for the potential, as a local potential is a function of $q$ only.

\begin{figure*}[tbh]
    \subfloat[]{\includegraphics[width=0.325\textwidth]{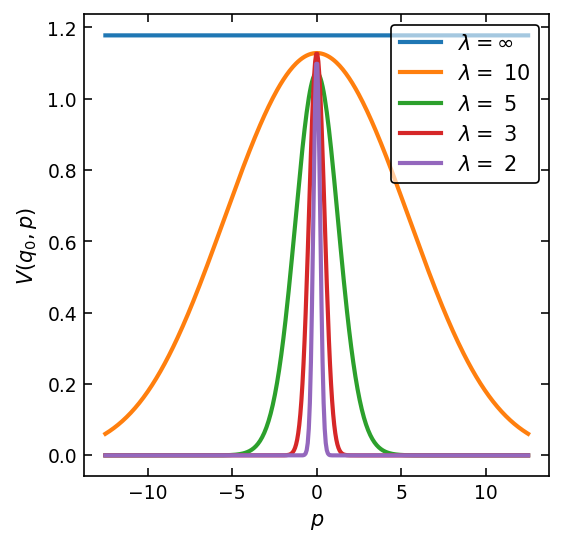}}%
    \subfloat[]{\includegraphics[width=0.32\textwidth]{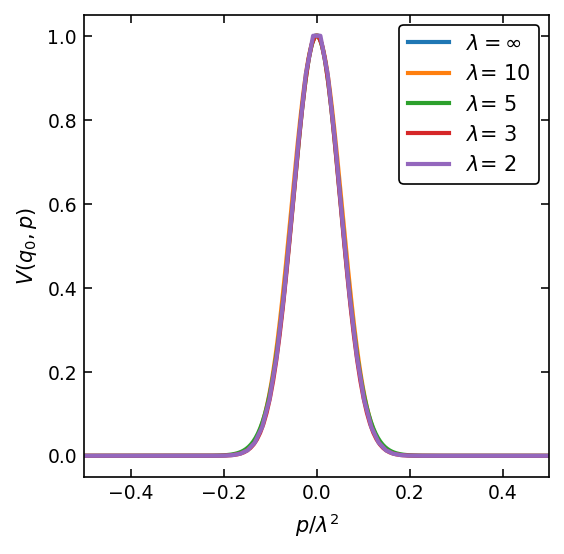}}%
    \subfloat[]{\includegraphics[width=0.333\textwidth]{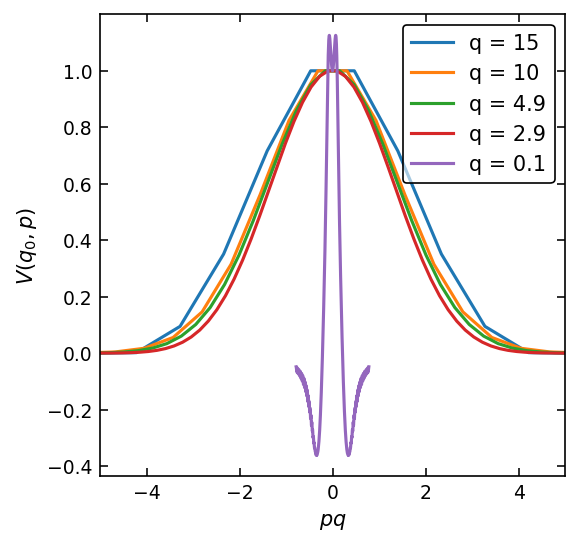}}%
    \caption{(a) SRG evolution of the one-level  two-Gaussian potential at fixed $q = q_0 = 7$.
    The original unevolved potential ($\lambda = \infty$) is fully local, as indicated by independence of $p$, however the SRG evolution progressively introduces nonlocality to the potential. 
    (b) Same as (a) but rescaled by $1/\lambda^2$, showing the dependence is only on $p/\lambda^2$ as implied by Eq.~\eqref{eq:SRG_nonlocality}.
    (c) Same potential but plotted against $pq_0$ for fixed $\lambda = 3$. 
    }
    \label{fig:Nonlocality_1} 
\end{figure*}

\begin{figure*}[tbh]
    \subfloat[]{\includegraphics[width=0.325\textwidth]{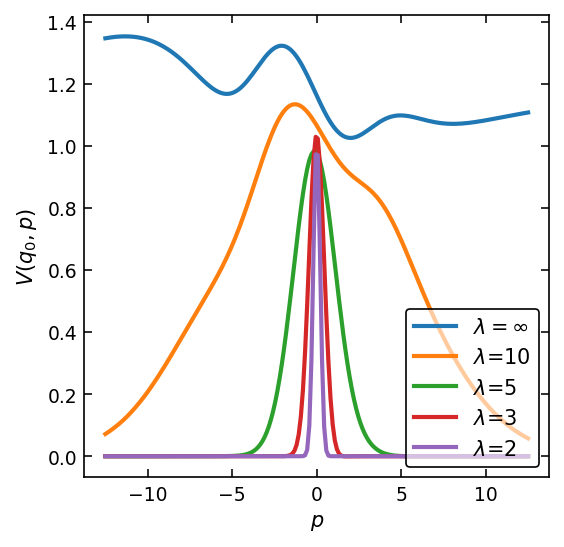}}%
    \subfloat[]{\includegraphics[width=0.32\textwidth]{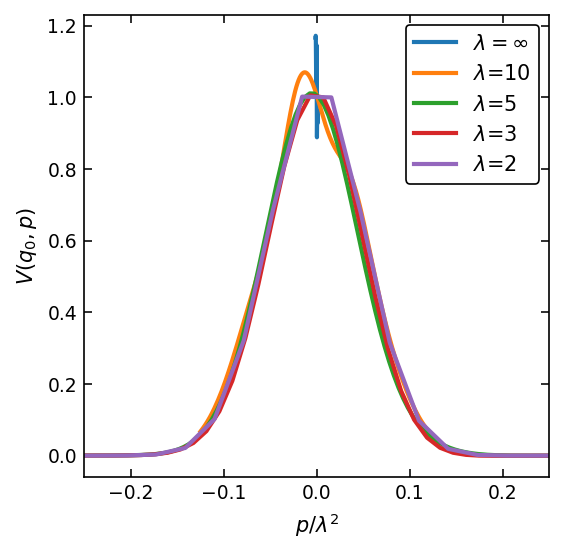}}%
    \subfloat[]{\includegraphics[width=0.336\textwidth]{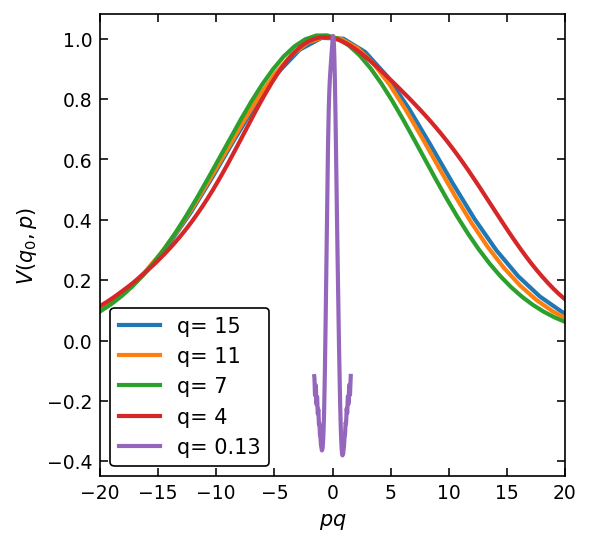}}%
    \caption{Same as Fig.~\ref{fig:Nonlocality_1} but for the SRG evolution of the optical potential derived from the two-level two-Gaussian potential. All the curves in panels (a) and (b) are for $q_0$ = 7.} 
    \label{fig:Nonlocality_2}
\end{figure*}

We can anticipate the induced nonlocality due to SRG evolution by focusing on the far off-diagonal elements in momentum space.
For nuclear-like Hamiltonians (including the one-dimensional analogs considered here), the diagonal is dominated by the kinetic energy, which leads to the SRG equation \eqref{eq:flow_eq} for these elements being given to good approximation by~\cite{Furnstahl:2012fn}
\begin{equation}
    \frac{d}{ds}V_s(k,k') = 
    -(k^2 - k'{}^2)^2 V_s(k,k').
\end{equation}
Thus each matrix element evolves independently and
the solution is (recalling $s = 1/\lambda^4)$
\begin{align}
 V_s(k,k') &= V_{s=0}(k,k')\,e^{-((k^2-k'{}^2)/
        \lambda^2)^2} \label{eq:SRG_offdiag}\\
        &= V_{s=0}(k,k')\, e^{-4p^2 q^2/\lambda^4}.
    \label{eq:SRG_nonlocality}
\end{align}
The first equality \eqref{eq:SRG_offdiag} manifests the suppression of off-diagonal matrix elements that decouple momenta separated by order $\lambda$ or greater.
The second equality \eqref{eq:SRG_nonlocality} manifests the $q$ and $p$ dependence that modifies the original potential.

If the initial potential is local, hence a function of q only, then from \eqref{eq:SRG_nonlocality} the nonlocality takes the form of the Gaussian dependence on $p$ with coefficient proportional to $q^2/\lambda^4$.
This in turn is proportional to the nonlocality width in coordinate space.
Thus the nonlocality extent increases rapidly as $\lambda$ decreases, and grows with increasing $q^2$. 
This behavior can be contrasted with Perey-Buck parametrized nonlocality~\cite{PEREY1962353} that is also Gaussian in $p$ but with a constant coefficient.
Note that these features are independent of the dimension.

We can test whether Eq.~\eqref{eq:SRG_nonlocality} captures the essence of the nonlocality by plotting the evolved potential with appropriate choices of independent variables. 
We SRG-evolve separately the even ($V^{+}$) and odd ($V^{-}$) components of the potential, given by:
\begin{align}
        V^{+} = \frac{1}{2}(V(k,k') + V(k,-k')) ,
        \\
        V^{-} = \frac{1}{2}(V(k,k') - V(k,-k')) .
\end{align}
At any $s$ we then combine these components to obtain the potential $V_s$ and the corresponding optical potential $\Vopt$ in the full $(k,k')$ space.

We begin with SRG evolution of the one-level two-Gaussian potential, which is plotted three ways for different $\lambda$ values in Fig.~\ref{fig:Nonlocality_1}.
Panel~(a) is the potential as a function of $p$ for a single representative value of $q = q_0 = 7$.
The unevolved potential ($\lambda=\infty$) 
has no $p$-dependence, indicating it is local. With decreasing $\lambda$ the width in $p$ decreases, meaning an increase in spatial nonlocality.
To test whether the dependence of the evolved potentials on $\lambda$ and $q$ are given by Eq.~\eqref{eq:SRG_nonlocality}, we scale the $x$-axis in panels (b) and (c).
Choosing a single $q$ value in (b) and plotting against $p/\lambda^2$, the curves overlap almost perfectly.
In panel (c) we choose a single SRG resolution scale ($\lambda=2$ in this case) and plot versus $pq$, so the width is determined by the constant $\lambda$ only. The graphs lie close to each other except for $q<\lambda$. 
Thus the nonlocality is indeed characterized by Eq.~\eqref{eq:SRG_nonlocality}.

Having set a baseline of nonlocal behavior for ordinary potentials, we turn to the optical potential for the two-level two-Gaussian potential in Fig.~\ref{fig:Nonlocality_2}.
In panel (a) at $\lambda = \infty$ and fixed $q = 7$,
we now observe the explicit nonlocality of the optical potential when unevolved. 
As we go lower in RG resolution, the nonlocality of the optical potential becomes overtaken by that of the SRG evolution.
This is manifest in panel (b) when we plot versus $p\lambda^2$, where the curves lie on top of each other once $\lambda < q_0$.
In panel (c) at constant $\lambda = 5$, we see the same dependence induced nonlocality in $q$ until $q < \lambda$.
Thus the SRG nonlocality dominates at low RG resolution with a ``universal'' Gaussian-like nonlocality proportional to $q$ and inversely proportional to $\lambda^2$.

\section{Summary and outlook}\label{sec:summary}

In this paper, we used a one-dimensional toy model to carry out exploratory studies of how an optical potential evolves under SRG evolution.
We showed that key features of the evolution of the full system to low resolution are inherited by the optical potential description.
In particular, high-momentum and low-momentum modes of the optical potential are decoupled while the observables are unchanged, and this is reflected in high-momentum components in bound states being suppressed and local decoupling of scattering states.
We also verified the increase in perturbativeness as measured by the repulsive Weinberg eigenvalues.
It followed that perturbative approximations for optical potentials, such as used in some ab initio approaches~\cite{Holt:2013tna}, become increasingly valid as the resolution is decreased.
We observed that the nonlocality of the optical potential gives way to that of the SRG evolution for momentum transfers greater than that of the SRG resolution scale $\lambda$.
Despite the simplicity of the toy model, due to the nature of the SRG via the flow equation, the observations in this study will likely carry over directly to three-dimensional examples.

In moving to such examples, a good candidate for the first system to consider is neutron-alpha particle scattering at different RG resolutions.
As the $n-\alpha$ optical potential describes one of the simplest processes of scattering off a composite particle, it is a natural 
continuation from the toy models explored here. We can compare the results of the $n-\alpha$ potential to those of current ab-initio calculations as well~\cite{Rotureau:2016jpf}.
Also, by extending the investigations in~\cite{More:2015tpa} in deuteron electrodisintegration, we can look at the corresponding problem with the breakup done by a neutron instead of an electron to better understand the theory/experiment discrepancy in scattering cross sections for breakup reactions~\cite{Tostevin:2014usa,Tostevin:2021ueu}.
In these studies we will
explore the feasibility of directly SRG evolving the optical potential.
This would require either evolving each of the ingredients in the equation for the optical potential \eqref{eq:feshbach_opt_pot} or working with the irreducible self-energy~\cite{Dickhoff:2018wdd}.
It will also be useful to consider alternative SRG schemes, such as using a generator for block decoupling (see Sec.~\ref{subsec:SRG}).

Because we have shown that basic features of RG evolution are inherited by optical potentials, we should also expect that the lessons from SRG evolution about the consistency of structure and reactions~\cite{Tropiano:2021qgf} will also apply.
In particular, the softening of low-energy many-body states, consistent with their description using shell model wave functions, is accompanied by modifications to reaction operators.
For electron scattering from nuclei, short-distance physics described at high-resolution with one-body operators is shifted to simple two-body operators~\cite{More:2017syr,Tropiano:2021qgf,Tropiano:2022jjj}.
This may mean significant three-body operator contributions for short-distance physics in nucleon-nucleus scattering.
In this regard, applying an RG analysis within the framework of the dispersion optical model (DOM)~\cite{Mahaux1991,Dickhoff:2016ikd,Dickhoff:2018wdd} and its recent applications~\cite{Dussan:2014vta,Atkinson:2019bwd} may prove fruitful.


\begin{acknowledgments}
We thank Scott Bogner for fruitful discussions and feedback.
This work was supported by the National Science Foundation under Grant No.~PHY--1913069, and the NUCLEI SciDAC Collaboration under US Department of Energy MSU subcontract RC107839-OSU\@.

\end{acknowledgments}

\bibliography{tropiano_bib}

\clearpage

\appendix*
\section{Supplemental Material}
\label{ap:supplemental}

In this Supplemental Material we present  figures for odd-parity states corresponding to selected figures for even-parity states in the main text.

\begin{figure*}[tbh]
     \subfloat[]{\includegraphics[width=0.45\textwidth]{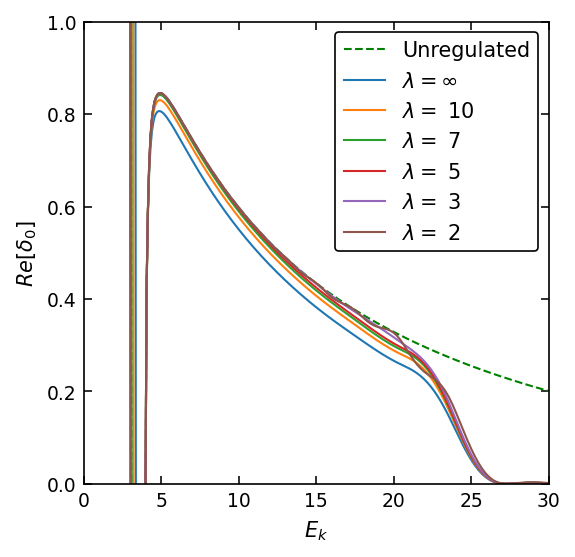}}
    \hspace{0.5cm}
    \subfloat[]{\includegraphics[width=0.46\textwidth]{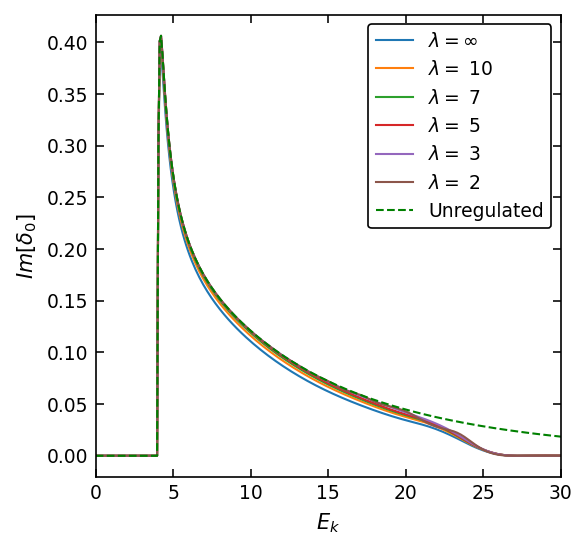}}
    \caption{Same as Fig~\ref{fig:Regulator_1}, but for the odd two-level two-Gaussian potential. The decoupling is not as apparent as that for the even-parity case due to less high-momentum strength. This is because the wave function passes through zero at the origin in coordinate space, so the odd states feel less of the repulsive part of the potential.} 
    \label{fig:odd_Regulator_1}
\end{figure*}

\begin{figure*}[tbh]
    \subfloat[]{\includegraphics[width=0.46\textwidth]{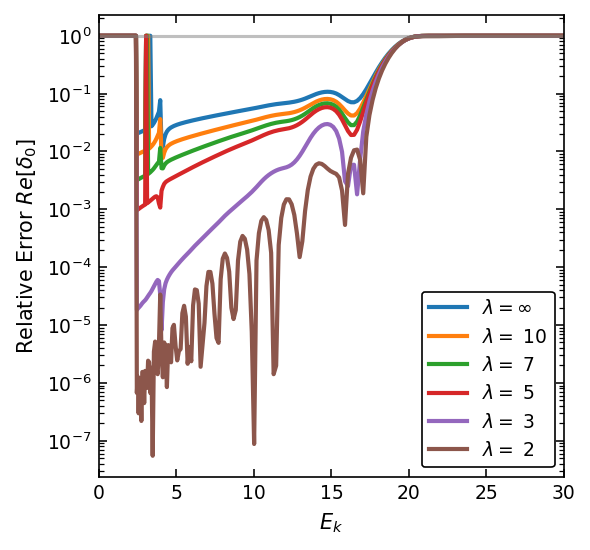}}
    \hspace{0.5cm}
    \subfloat[]{\includegraphics[width=0.46\textwidth]{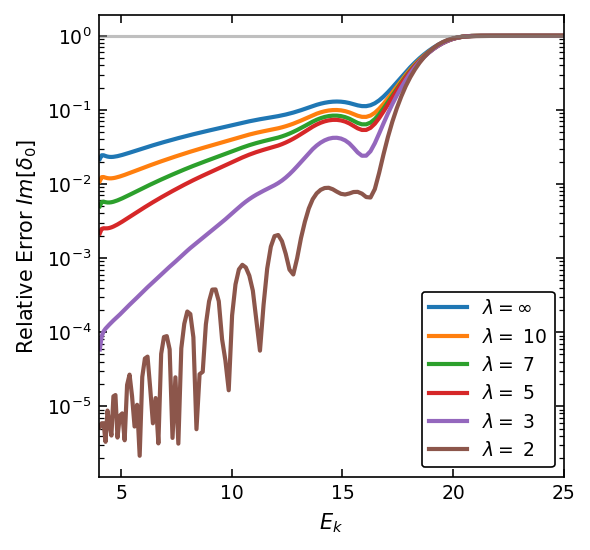}}
    \caption{Same as Fig.~\ref{fig:odd_Regulator_1} but plotting the relative errors of the odd-parity phase shifts from regulated SRG-evolved optical potentials compared to the exact phase shifts.
    The light gray horizontal line marks 100\% relative error.} 
    \label{fig:odd_Regulator_2}
\end{figure*}

\begin{figure*}[tbh!]
    \subfloat[]{\includegraphics[width=0.33\textwidth]{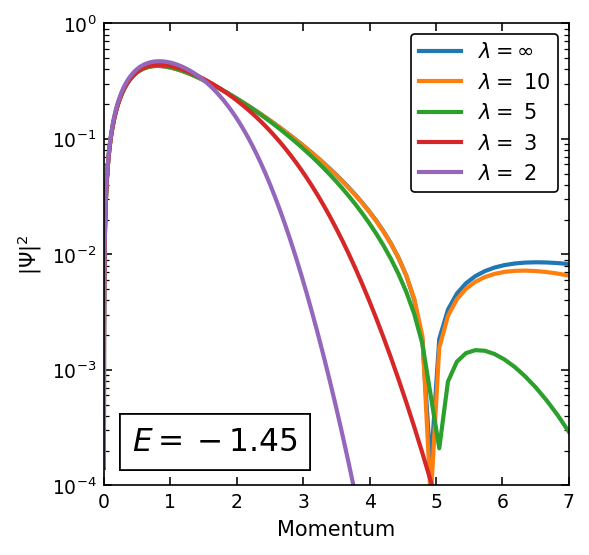}}%
    \subfloat[]{\includegraphics[width=0.33\textwidth]{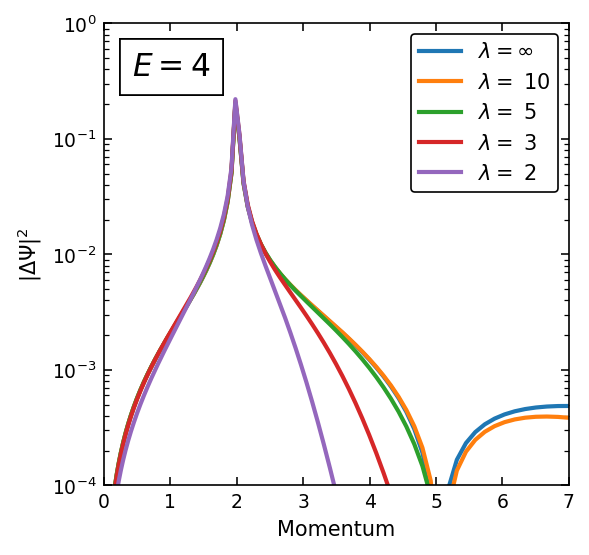}}%
    \subfloat[]{\includegraphics[width=0.33\textwidth]{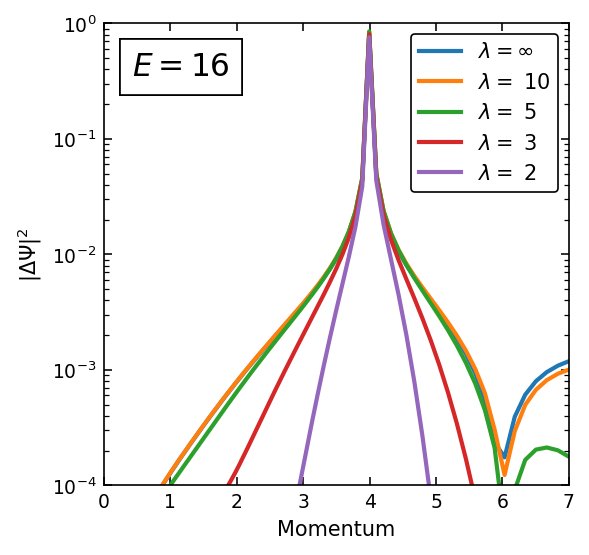}}
    \caption{Odd-parity wave functions as in Fig.~\ref{fig:wf_1} from using the odd two-level two-Gaussian potential. They display the same pattern of decoupling as did the even-parity wave functions.
    However, the high-momentum strength is less than for the even-parity states because the effect of the repulsion is less as the odd states go through zero at the origin.} 
    \label{fig:wf_1_odd}
\end{figure*}

\end{document}